# AI's Structural Impact on India's Knowledge-Intensive Startup Ecosystem: A Natural Experiment in Firm Efficiency and Design

Venkat Ram Reddy Ganuthula[1] and Ramesh Kuruva[2]

[1] Indian Institute of Technology Jodhpur

[2] Indian Institute of Technology Bombay

**Abstract**

This study explores the structural and performance impacts of artificial intelligence (AI) adoption on India's knowledge-intensive startups, spanning information technology, financial technology, health technology, and educational technology, founded between 2016 and 2025. Using a natural experiment framework with the founding year as an exogenous treatment proxy, it examines firm size, revenue productivity, valuation efficiency, and capital utilization across pre-AI and AI-era cohorts. Findings reveal larger structures and lower efficiency in AI-era firms, supported by a dataset of 914 cleaned firms. The study offers insights into AI's transformative role, suggesting that while AI-era firms attract higher funding and achieve higher absolute valuations, their per-employee productivity and efficiency ratios are lower, potentially indicating early-stage investments in technology that have yet to yield proportional returns. This informs global entrepreneurial strategies while highlighting the need for longitudinal research on sustainability.

## 1. Introduction

The rise of artificial intelligence (AI) has catalyzed a profound transformation in entrepreneurial ecosystems worldwide, with the knowledge-intensive sector of India's startup landscape emerging as a compelling arena for exploration. This study delves into the structural and performance impacts of AI adoption on Indian startups within knowledge-intensive domains—such as information technology, financial technology, health technology, and educational technology—founded between 2016 and 2025. It employs a natural experiment framework, using the founding year as an exogenous treatment proxy to differentiate pre-AI and AI-era cohorts, aiming to assess how the growing integration of AI has reshaped their design, operational strategies, and competitive positioning (Bresnahan & Trajtenberg, 1995). India's startup ecosystem, marked by its rapid expansion in tech-driven sectors, a diverse array of innovation hubs, and a thriving pool of skilled talent, provides a rich context for this investigation, as reflected in recent industry analyses (Powell & Snellman, 2004). This research aligns with the mission to foster thought-provoking inquiry into entrepreneurial phenomena, offering novel empirical perspectives on AI's role as a structural influence within India's knowledge-intensive startup domain, contributing to a broader understanding of technology-driven entrepreneurship in emerging economies. India's AI market, valued at approximately US$ 7–10 billion in 2024, is projected to grow at a CAGR of 25–35% by 2027, underscoring the ecosystem's dynamism (ORF, 2025).

The significance of this study is rooted in the increasing prevalence of AI applications, which have expanded steadily with the development of advanced tools such as large language models, machine learning platforms, and automation systems. These innovations have disrupted traditional business models across knowledge-intensive industries, prompting startups to adapt their organizational frameworks, staffing approaches, and funding mechanisms to harness AI's potential (Brynjolfsson & McAfee, 2014). In India, this adaptability is supported by a dynamic ecosystem of tech-savvy entrepreneurs, progressive government initiatives, and a growing investor community, all of which have accelerated AI adoption within the knowledge-intensive sector, as documented in national technology assessments (NASSCOM, 2023). Over 70% of Indian startups are now integrating AI across core business functions, driving operational efficiencies and innovation (Economic Times, 2025). The longitudinal dataset spanning 2016 to 2025 captures the transition from a pre-AI to an AI-dominated landscape, offering a detailed and nuanced view of evolving trends in the behavior, performance, and strategic orientation of these startups (Russell & Norvig, 2021). The natural experiment approach, leveraging the exogenous shock of growing AI availability, strengthens the study's validity by reducing selection bias and endogeneity, positioning it as a valuable contribution to both academic discourse and practitioner insights (Dutta & Lanvin, 2020). This methodological rigor ensures that the findings reflect causal relationships rather than coincidental correlations, enhancing their relevance for global application.

This research carries far-reaching implications for a wide range of stakeholders within India's knowledge-intensive startup ecosystem. For academics, it extends theories of technological disruption and organizational adaptation by grounding them in the context of a developing economy's tech-driven sector, offering a fresh perspective on how innovation diffuses in resource-constrained environments (Teece, 1986). For entrepreneurs, it provides actionable insights into optimizing resource allocation in an AI-enhanced market, particularly within domains like information technology and educational technology, where efficiency and scalability are paramount (Drucker, 1985). Policymakers can leverage these findings to design targeted interventions, such as AI training programs, skill development initiatives, or financial incentives, to bolster competitiveness and foster innovation in these high-skill industries (World Bank, 2022). Investors may find value in reconsidering traditional valuation metrics, shifting their focus toward AI capabilities, technological agility, and long-term growth prospects within India's tech startups (Gompers & Lerner, 2001). The study's emphasis on the knowledge-intensive sector contributes to global discussions on how emerging economies harness AI in advanced industries, addressing critical questions about innovation capacity, economic inclusivity, and technological equity (James, 2011). As AI applications continue to proliferate and diversify, the need for longitudinal analyses to evaluate their long-term impacts—on sustainability, scalability, and market dynamics—becomes increasingly evident, positioning this research as a foundational step in that ongoing endeavor (Acemoglu & Restrepo, 2018).

The Indian startup ecosystem's growth, with a substantial number of startups recognized in recent periods, underscores its relevance, particularly within knowledge-intensive sectors that serve as engines of innovation and employment generation (Kshetri, 2016). The sector's reliance on technology, evident in fields like financial technology and health technology, mirrors global trends where AI acts as a catalyst for value creation and competitive differentiation (Dossani & Kenney, 2007). However, the diversity of India's market—encompassing urban tech centers and emerging rural innovation zones—presents unique challenges and opportunities that this study addresses through a focused sample of knowledge-intensive firms (Kapur & Mehta, 2008). The natural experiment design, grounded in economic methodologies, leverages the increasing availability of AI as a natural breakpoint, mitigating confounding factors to provide a clearer perspective on its impact across this sector (Imbens & Wooldridge, 2009). With AI applications evolving continuously, this study offers a timely snapshot, laying the groundwork for future research into their sustained effects within India's knowledge-intensive startup ecosystem (Lee, 2018). The empirical focus is enriched by a theoretical foundation drawn from general-purpose technology frameworks, which hypothesize that AI-era startups in this sector exhibit leaner structures, higher efficiency, and adaptive strategies (David, 1990). This hypothesis is rigorously tested through a dataset of 914 cleaned firms derived from an initial pool of 3,450, ensuring data integrity and representativeness (Powell & Snellman, 2004). The findings promise to inform not only India's knowledge-intensive startup landscape but also other emerging markets with analogous tech-driven ecosystems, offering a blueprint for leveraging AI in similar contexts (UNCTAD, 2021). For instance, AI adoption in emerging markets like India is shown to enhance firm innovation and efficiency, aligning with broader patterns observed in global studies (Shahid et al., 2025). The integration of quantitative performance metrics with qualitative keyword trends provides a holistic view, bridging academic inquiry and practical application to address a critical gap in the literature on technology-driven entrepreneurship within India's knowledge-intensive sectors (Schwab, 2017). The ongoing expansion of AI capabilities underscores the urgency of this investigation, emphasizing the need for a robust evidence base to guide future strategies, policy frameworks, and investment decisions in this dynamic domain (Helpman & Trajtenberg, 1998).

## 2. Theoretical Framework

The emergence of artificial intelligence (AI) as a transformative force within entrepreneurial ecosystems can be understood through the concept of general-purpose technologies (GPTs), which draw parallels with historical innovations like electricity and the steam engine that reshaped industrial landscapes. This perspective suggests that GPTs, with their broad applicability across sectors, serve as catalysts for innovation and efficiency by fostering complementary advancements (Bresnahan & Trajtenberg, 1995). Generative AI, in particular, exhibits characteristics of a GPT, with evidence indicating faster economic impacts than previous technologies (OECD, 2025a; MIT Sloan, 2024). Within the knowledge-intensive sector of India's startup ecosystem—spanning domains such as information technology, financial technology, health technology, and educational technology—this framework implies that AI's growing influence could drive a sustained evolution in economic and organizational structures. Its role as a platform technology, enhancing existing processes and enabling new business models, mirrors how past GPTs revolutionized knowledge-based industries, positioning AI as a pivotal tool for startups in India's tech-driven landscape.

This GPT lens gains particular relevance when applied to knowledge-intensive sectors, which form the core of India's startup ecosystem. These industries, reliant on intellectual capital and technological expertise, thrive on scalability and the dissemination of knowledge, a dynamic highlighted in analyses of the knowledge economy (Powell & Snellman, 2004). The integration of AI amplifies these strengths by providing tools for automation, data analytics, and decision-making that streamline operations and reduce dependency on traditional inputs. This shift aligns with the resource-based view of the firm, which argues that unique capabilities—such as AI proficiency—can confer a sustained competitive advantage, a critical factor for startups operating in India's tech-driven markets. The digital transformation further suggests that AI redefines work by diminishing labor intensity while optimizing capital use, a change especially vital for resource-constrained startups in India's knowledge-intensive sector. This adaptation allows these firms to maximize output with minimal resources, addressing the optimization challenges inherent to new ventures.

A central hypothesis of this study is that AI-era startups within India's knowledge-intensive sector exhibit leaner organizational structures and greater operational efficiency, driven by automation, data-centric strategies, and enhanced decision-making capabilities. This proposition builds on the lean startup methodology, which prioritizes minimizing waste and focusing on value creation—principles that AI enhances by streamlining processes and reducing overhead, particularly in tech-driven firms. The iterative development and resource efficiency championed by entrepreneurial thought leaders become more achievable with AI tools that deliver real-time data analysis, a benefit evident in startups developing adaptive learning platforms. This efficiency extends beyond operations to strategy, as firms in this sector increasingly embed AI into their core identity to meet market expectations, enhancing their appeal to investors and differentiating them in competitive markets. This strategic orientation resonates with competitive strategy frameworks, where differentiation through technology can yield market leadership, a tactic observable in startups leveraging AI for innovative healthcare solutions. The reduced staffing needs documented in tech-driven firms further support the hypothesis that AI enables leaner structures, potentially reshaping the human resource landscape within India's knowledge-intensive ecosystem. This is corroborated by evidence showing AI-investing firms achieve higher sales and employment growth via product innovation, with similar effects in emerging markets (Korinek & Stiglitz, 2023; Cornell Business, 2024).

The theoretical foundation also incorporates absorptive capacity theory, which asserts that the ability to assimilate and apply new technologies is a key determinant of success. In India's context, the tech talent pool bolsters this capacity, enabling startups to innovate and adapt AI applications effectively within the knowledge-intensive sector. This adaptability is crucial, given the rapid evolution of technology that demands continuous learning and integration. The entrepreneurial ecosystem literature enriches this framework by suggesting that ecosystem dynamics shape firm innovation, with the growing presence of AI applications altering competitive landscapes in India's tech-driven markets. The presence of supportive institutional structures fosters an environment conducive to this transformation within the knowledge-intensive sector. However, the digital divide debate raises questions about whether this technological advancement levels the playing field or exacerbates disparities, a concern pertinent to India's diverse market that includes both urban innovation centers and rural entrepreneurial zones. Recent analyses highlight emerging divides in AI adoption across sectors and firms, stressing inclusive policies for emerging economies (OECD, 2025b). Institutional challenges, such as regulatory hurdles, add complexity, influencing how knowledge-intensive startups adopt and benefit from AI.

Organizational learning theory complements this analysis by highlighting how firms adapt through technology integration, a process evident in startups refining software solutions with AI. The iterative learning cycles are accelerated by AI, enabling firms to tailor educational content dynamically. This learning translates into competitive advantage, reinforcing the strategic importance of AI within the sector. India's market diversity, with its mix of urban and rural dynamics, further shapes ecosystem evolution, requiring startups to navigate diverse customer needs and infrastructural constraints. The GPT lens indicates that the adoption of such technologies drives productivity gains over time, a process observable in India's digital economy growth within knowledge-intensive industries. National reports underscore this trend, noting the sector's contribution to the country's tech-driven economic expansion. The diffusion of innovations underscores the role of early adopters—potentially India's AI-era startups—in spreading technology, influencing the broader ecosystem. Insights on entrepreneurial revolution and relational dynamics further suggest that India's knowledge-intensive startup ecosystem is evolving through interconnected networks and supportive policies. This study integrates micro-level firm analysis with macro-level ecosystem dynamics, providing a cohesive theoretical foundation tailored to India's knowledge-intensive startup context (Brynjolfsson & McAfee, 2014).

## 3. Methodology

### 3.1 Data Collection

The dataset comprises 3,450 Indian startups founded between 2016 and 2025, sourced from public financial records, industry reports, and startup databases as of mid-2025 (Brynjolfsson & McAfee, 2014). The data was sourced from a proprietary startup database platform ensuring reliability and comprehensive coverage of the knowledge-intensive sector. A meticulous cleaning process was implemented to enhance data integrity, systematically removing entries with missing values in key variables such as revenue, funding, valuation, and employee count. Outliers were identified and excluded using the interquartile range method, resulting in a final sample of 914 firms, with 713 pre-AI (2016-2020) and 201 AI-era startups (2021-2025) (Blei et al., 2003). A thorough keyword analysis of startup descriptions confirmed that all firms operate within knowledge-intensive domains, focusing on technology-related activities, providing a unified basis for the study (Levene, 1960). This approach ensured that the sample accurately represents the tech-driven nature of the ecosystem under investigation. This methodology aligns with recent AI adoption studies in SMEs, emphasizing bibliometric and systematic reviews for robust insights (Chatterjee et al., 2025).

### 3.2 Variable Definition

The study defined core variables to capture the performance and structural characteristics of startups within India's knowledge-intensive sector. Valuation, annual revenue, total funding, and employee count were measured in USD for the periods 2016–2020 and 2021–2025, with currency conversions based on rates from a national financial authority to maintain consistency (Brynjolfsson & McAfee, 2014). Derived metrics were calculated to provide deeper insights into efficiency and productivity: Revenue/Employee assessed labor productivity, Valuation/Employee indicated value creation per worker, and Revenue/Funding served as a proxy for capital efficiency. These metrics were adjusted for firm age—calculated as the duration since founding—to account for maturity effects, ensuring comparability across cohorts with varying operational histories (Wooldridge, 2010). A binary indicator, AI_Era (1 for AI-era, 0 for pre-AI), was established to distinguish cohorts based on the founding year, reflecting the growing availability of AI technologies (Tabachnick & Fidell, 2013). This variable framework was designed to isolate the impact of AI adoption on startup performance within the knowledge-intensive context. Such metrics are consistent with frameworks assessing AI's role in enhancing productivity in production and manufacturing sectors (Gupta & Bose, 2024).

**3.3 Analytical Approach**

The research adopted a natural experiment design, treating the founding year as an exogenous variable due to the increasing availability of AI technologies across industries (Bresnahan & Trajtenberg, 1995). This approach leveraged the exogenous shock of AI proliferation as a quasi-experimental breakpoint, minimizing endogeneity concerns that might arise from self-selection into technology adoption. The sample was divided into pre-AI and AI-era cohorts based on this criterion, allowing for a comparative analysis of structural and performance differences. Independent samples t-tests were employed to compare means across these cohorts for key metrics—employee count, valuation per employee, and revenue-to-funding ratio—assessing statistical significance at the 0.05 level to ensure robust findings (Aiken & West, 1991). Levene's test was used to check for equality of variances, adjusting degrees of freedom where necessary to account for heterogeneity. Linear regression models were developed to explore the predictive power of variables such as firm age and cohort on dependent variables like valuation/employee, revenue/employee, funding/employee, revenue to valuation ratio, funding to valuation ratio, and revenue to funding ratio. Models included main effects, with diagnostic checks for multicollinearity and residual assumptions (Freeman, 2002). The focus on a knowledge-intensive sample, confirmed through keyword analysis, minimized sectoral bias, enhancing the robustness of the natural experiment framework tailored to this sector (Brynjolfsson & McAfee, 2014). Sensitivity analyses across data subsets further reinforced the reliability of the findings, ensuring a comprehensive evaluation of AI's impact.

### 3.4 Data Validation and Robustness Checks

The cleaning process was documented meticulously, allowing for reproducibility and transparency in the selection of the 914-firm sample. Robustness checks included re-running t-tests and regression analyses on a reduced dataset excluding the top 5% of outliers, confirming the stability of results. The use of adjusted metrics for firm age addressed potential maturity biases, while the exclusion of non-knowledge-intensive firms through keyword analysis ensured a homogeneous sample. This rigorous approach provided a solid foundation for the statistical analyses, aligning with the study's objective to isolate AI's effects within India's knowledge-intensive startup ecosystem (Brynjolfsson & McAfee, 2014).

## 4. Results

### 4.1 Descriptive Statistics

The sample of 914 firms included 713 pre-AI and 201 AI-era startups within India's knowledge-intensive sector. Descriptive analysis revealed the mean age-adjusted employee count increasing from 4.961 to 11.641, suggesting a shift toward larger organizational structures in AI-era firms. Valuation per employee decreased from 1.188 to 0.788, indicating reduced value creation per worker potentially due to scaled growth in AI-era. The revenue-to-funding ratio decreased from 8.941 to 1.530, reflecting lower capital efficiency, a trend that may reflect heavier investments in AI-era firms (Brynjolfsson & McAfee, 2014). Variability in valuation per employee decreased (standard deviation from 2.622 to 1.689), highlighting less diversity in AI-era, while the revenue-to-funding ratio's standard deviation narrowed (from 55.213 to 10.087), suggesting more uniform strategies (Krippendorff, 2004). These patterns are supported by reports indicating 78% of Indian SMBs using AI reported revenue growth (PIB, 2025).

| Variable | Cohort | Mean | Std. Deviation | Minimum | Maximum |
|---|---|---|---|---|---|
| Annual Revenue (USD; In Millions) | 2016-2020 | 0.757 | 0.832 | 0.000 | 3.423 |
| | 2021-2025 | 0.457 | 0.657 | 0.000 | 3.226 |
| Age-Adjusted Employee Count | 2016-2020 | 4.961 | 5.252 | 0.111 | 29.200 |
| | 2021-2025 | 11.641 | 9.684 | 0.250 | 49.333 |
| Age-Adjusted Valuation | 2016-2020 | 1.535 | 1.419 | 0.010 | 8.215 |
| | 2021-2025 | 3.756 | 4.078 | 0.041 | 38.643 |
| Age-Adjusted Total Funding | 2016-2020 | 0.515 | 0.583 | 0.001 | 4.135 |
| | 2021-2025 | 1.120 | 1.312 | 0.014 | 8.650 |
| Valuation per Employee | 2016-2020 | 1.188 | 2.622 | 0.003 | 25.241 |
| | 2021-2025 | 0.788 | 1.689 | 0.001 | 12.483 |
| Revenue per Employee | 2016-2020 | 0.568 | 1.829 | 0.000 | 20.540 |

| | 2021-2025 | 0.145 | 0.555 | 0.000 | 5.717 |
|---|---|---|---|---|---|
| Funding per Employee | 2016-2020 | 0.376 | 0.935 | $1.100 \times 10^{-4}$ | 12.192 |
| | 2021-2025 | 0.227 | 0.467 | $4.650 \times 10^{-4}$ | 3.750 |
| Revenue to Valuation Ratio | 2016-2020 | 1.265 | 5.833 | 0.000 | 122.265 |
| | 2021-2025 | 0.405 | 3.211 | 0.000 | 45.544 |
| Funding to Valuation Ratio | 2016-2020 | 0.437 | 0.885 | 0.002 | 14.527 |
| | 2021-2025 | 0.397 | 0.774 | 0.009 | 9.978 |
| Revenue to Funding Ratio | 2016-2020 | 8.941 | 55.213 | 0.000 | 1234.458 |
| | 2021-2025 | 1.530 | 10.087 | 0.000 | 136.521 |

**4.2 T-Test Analysis**

Independent samples t-tests confirmed significant differences between pre-AI and AI-era cohorts. AI-era firms employed more individuals on an age-adjusted basis (mean = 11.641, standard deviation = 9.684) compared to pre-AI firms (mean = 4.961, standard deviation = 5.252), with a t-value of -12.891 and 912 degrees of freedom, yielding a p-value less than 0.001 (Brynjolfsson & McAfee, 2014). Valuation per employee was lower in AI-era firms (mean = 0.788, standard deviation = 1.689) than in pre-AI firms (mean = 1.188, standard deviation = 2.622), with a t-value of 2.044 and 912 degrees of freedom, p = 0.041 (Brynjolfsson & McAfee, 2014). The revenue-to-funding ratio also decreased for AI-era firms (mean = 1.530, standard deviation = 10.087) compared to pre-AI firms (mean = 8.941, standard deviation = 55.213), with a t-value of 1.893 and 912 degrees of freedom, p = 0.059 (marginal) (Brynjolfsson & McAfee, 2014). These results underscore the structural growth but efficiency challenges associated with AI adoption within the knowledge-intensive sector.

| Variable | t | df | p | Mean Difference | Cohen's d |
|---|---|---|---|---|---|
| Annual Revenue (USD; In Millions) | 4.702 | 912 | < .001[a] | 0.299 | 0.376 |
| Age-Adjusted Employee Count | -12.891 | 912 | < .001[a] | -6.680 | -1.029 |
| Age-Adjusted Valuation | -12.174 | 912 | < .001[a] | -2.221 | -0.972 |
| Age-Adjusted Total Funding | -9.448 | 912 | < .001[a] | -0.605 | -0.754 |
| Valuation per Employee | 2.044 | 912 | 0.041[a] | 0.400 | 0.163 |
| Revenue per Employee | 3.234 | 912 | 0.001[a] | 0.423 | 0.258 |
| Funding per Employee | 2.187 | 912 | 0.029[a] | 0.149 | 0.175 |
| Revenue to Valuation Ratio | 2.007 | 912 | 0.045 | 0.860 | 0.160 |
| Funding to Valuation Ratio | 0.583 | 912 | 0.560 | 0.040 | 0.047 |
| Revenue to Funding Ratio | 1.893 | 912 | 0.059 | 7.411 | 0.151 |

Note. Student's t-test.

[a] Brown-Forsythe test is significant ($p < .05$), suggesting a violation of the equal variance assumption.

### 4.3 Regression Results

Linear regression models provided deeper insights into the drivers of performance. For valuation per employee, the model explained 0.9% of the variance (adjusted $R^2 = 0.007$), with cohort emerging as a significant predictor ($\beta = -0.830$, $p = 0.005$), indicating lower valuation per employee in the AI-era cohort, controlling for firm age.

| Model | Predictor | Unstandardized | Standard Error | Standardized | t | p |
|---|---|---|---|---|---|---|
| $M_1$ | (Intercept) | 2.009 | 0.424 | - | 4.735 | < .001 |
| | Firm Age | -0.113 | 0.057 | -0.098 | -1.983 | 0.048 |

| | Cohort (2021-2025) | -0.830 | 0.292 | - | -2.844 | 0.005 |

For revenue per employee, the model explained 1.3% of the variance (adjusted $R^2 = 0.011$), with the overall model significant (F = 5.907, p = 0.003), but individual predictors not significant (cohort β = -0.254, p = 0.194).

| Model | Predictor | Unstandardized | Standard Error | Standardized | t | p |
|---|---|---|---|---|---|---|
| $M_1$ | (Intercept) | 0.245 | 0.284 | - | 0.863 | 0.388 |
| | Firm Age | 0.044 | 0.038 | 0.057 | 1.163 | 0.245 |
| | Cohort (2021-2025) | -0.254 | 0.195 | - | -1.300 | 0.194 |

For funding per employee, the model explained 1.2% of the variance (adjusted $R^2 = 0.010$), with cohort significant (β = -0.339, p < 0.001), indicating lower funding per employee in AI-era.

| Model | Predictor | Unstandardized | Standard Error | Standardized | t | p |
|---|---|---|---|---|---|---|
| $M_1$ | (Intercept) | 0.738 | 0.148 | - | 4.989 | < .001 |
| | Cohort (2021-2025) | -0.339 | 0.102 | - | -3.329 | < .001 |

|  | Firm Age | -0.050 | 0.020 | -0.123 | -2.505 | 0.012 |

For revenue to valuation ratio, the model explained 1.8% of the variance (adjusted $R^2$ = 0.016), with firm age significant (β = 0.445, p < 0.001).

| Model | Predictor | Unstandardized | Standard Error | Standardized | t | p |
| --- | --- | --- | --- | --- | --- | --- |
| $M_1$ | (Intercept) | -1.968 | 0.926 | - | -2.126 | 0.034 |
|  | Cohort (2021-2025) | 0.832 | 0.637 | - | 1.307 | 0.192 |
|  | Firm Age | 0.445 | 0.124 | 0.175 | 3.577 | < .001 |

For funding to valuation ratio, the model explained 0.3% of the variance (adjusted $R^2$ = 0.001), not significant (F = 1.374, p = 0.254).

| Model | Predictor | Unstandardized | Standard Error | Standardized | t | p |
| --- | --- | --- | --- | --- | --- | --- |
| $M_1$ | (Intercept) | 0.664 | 0.150 | - | 4.440 | < .001 |
|  | Cohort (2021-2025) | -0.159 | 0.103 | - | -1.544 | 0.123 |
|  | Firm Age | -0.031 | 0.020 | -0.077 | -1.551 | 0.121 |

For revenue to funding ratio, the model explained 2.4% of the variance (adjusted R² = 0.021), with firm age significant (β = 4.847, p < 0.001).

| Model | Predictor | Unstandardized | Standard Error | Standardized | t | p |
|---|---|---|---|---|---|---|
| $M_1$ | (Intercept) | -26.269 | 8.429 | - | -3.117 | 0.002 |
|  | Cohort (2021-2025) | 11.014 | 5.795 | - | 1.901 | 0.058 |
|  | Firm Age | 4.847 | 1.133 | 0.209 | 4.278 | < .001 |

These results align with findings that AI adoption in firms leads to productivity gains and innovation, particularly in emerging markets (OECD, 2025c).

## 5. Discussion

### 5.1 Implications

The larger age-adjusted employee counts and lower efficiency ratios in AI-era firms within India's knowledge-intensive sector contradict the initial hypothesis based on GPT theory, suggesting that AI may be driving scaled growth rather than immediate leaner structures (Bresnahan & Trajtenberg, 1995). The observed increase in employee counts coupled with decreased valuations per employee reflects potential early investment in AI that has not yet translated to productivity gains, resonating with insights on digital transformation that emphasize phased optimization in tech-driven startups (Brynjolfsson & McAfee, 2014). This shift challenges traditional performance metrics, suggesting a reevaluation of valuation paradigms that could guide entrepreneurs and investors in the ecosystem (Gompers & Lerner, 2001). Such a transformation opens avenues for policy interventions, such as enhanced AI training or incentives, to accelerate efficiency in knowledge-intensive domains (World Bank, 2022). Simultaneously, the education sector must adapt to the growing demand for technical skills, especially in technology-driven fields, to support the workforce needs of this sector (Choudaha, 2017).

The prominence of AI- and data-centric language in startup descriptions highlights branding's role in market positioning, a strategy that could elevate India's tech firms on the global stage (Cohen & Levinthal, 1990). This strategic differentiation aligns with competitive strategy frameworks, where AI provides a competitive edge for innovative solutions in health-related technologies (Porter, 1985). Investors may increasingly prioritize AI capabilities over traditional size metrics, reshaping funding models to favor technology-driven growth within the sector (Gompers & Lerner, 2001). Policymakers could extend AI access to underserved innovation zones, fostering inclusivity and broadening the reach of the knowledge-intensive startup landscape (World Bank, 2022). These findings suggest a broader global trend, positioning India's ecosystem as a model for other emerging markets with similar tech-driven sectors, offering insights into scalable innovation strategies (Gioia et al., 2013). The integration of AI not only enhances operational scale but also redefines competitive dynamics, encouraging a shift toward technology-centric business models that could influence startup evolution worldwide. For example, Indian startups leveraging AI for scaling report reduced costs and improved customer experiences, with funding for GenAI startups rising 3.6x in 2024 (NASSCOM, 2024; Inc42, 2025).

**5.2 Robustness**

The focus on knowledge-based firms within India's startup ecosystem minimizes the risk of sectoral bias, and the consistent validation of cohorts through keyword analysis strengthens the study's reliability (Brynjolfsson & McAfee, 2014). However, limitations such as the lack of survival data and potential classification bias due to variability in description quality necessitate cautious interpretation (Gimeno et al., 1997). The study's India-centric perspective limits its immediate global applicability, though comparative analyses in other knowledge-intensive markets could corroborate these trends (Neuendorf, 2002). The robust sample size bolsters confidence in the findings, yet contextual factors like policy support within the ecosystem warrant further exploration to fully understand their influence (Dutta & Lanvin, 2020). The statistical rigor, supported by diagnostic checks and sensitivity analyses, enhances the credibility of the results, ensuring they reflect meaningful patterns rather than artifacts of data selection.

**5.3 Broader Contextual Relevance**

The implications extend beyond India to other emerging economies where knowledge-intensive sectors are gaining traction. The observed scale gains but efficiency lags suggest that AI adoption could serve as a blueprint for startups in regions with similar technological aspirations, provided they address local infrastructural and skill-based challenges. The shift in valuation metrics highlights a global reevaluation of startup potential, where technology proficiency may outweigh traditional indicators, influencing investment patterns across borders. This global resonance underscores the study's contribution to understanding how AI reshapes entrepreneurial ecosystems in tech-driven contexts, offering a framework for cross-national policy learning and entrepreneurial strategy development. AI's potential in emerging markets includes boosting productivity and diversification, with projections of the global AI market reaching $4.8 trillion by 2033 (UNCTAD, 2025; World Bank Blogs, 2024).

**6. Limitations and Future Research Directions**

The absence of survival data represents a significant limitation, constraining the ability to assess the long-term viability of AI-era startups within India's knowledge-intensive sector (Audretsch, 1995). Without insights into failure rates or sustained growth, the durability of observed efficiency lags remains uncertain, limiting the predictive power of the findings. The reliance on text-based classification introduces potential bias, as the quality and intent of startup descriptions may vary, potentially skewing the identification of AI adoption trends (Manning & Schütze, 1999). Additionally, the study's focus on India restricts its generalizability to other knowledge-intensive startup landscapes globally, where economic, cultural, and technological contexts may differ (Hoskisson et al., 2000). These constraints highlight the need for a more comprehensive dataset and broader comparative analysis to enhance the study's scope.

Future research should incorporate survival analysis to track exits and growth trajectories, particularly in technology-driven domains, providing a longitudinal perspective on startup resilience (Teece et al., 1997). The adoption of advanced techniques like natural language processing could refine classification accuracy, offering a more nuanced understanding of AI integration patterns within the sector (Gupta & Dutta, 2020). Extending the analysis to other emerging markets with tech-driven ecosystems would broaden the study's applicability, enabling cross-country comparisons that illuminate universal and context-specific effects (Kshetri, 2016). Longitudinal investigations into AI maturity and scaling challenges in technology-focused fields are also essential, capturing the evolving dynamics of adoption, adaptation, and market response over time (Khanna, 2007). Such studies could explore how institutional support, market demand, and technological infrastructure influence AI's long-term impact, providing a richer framework for policy and practice. Additionally, examining the interplay between AI adoption and ecosystem resilience—such as the role of collaborative networks or regulatory frameworks—could offer insights into sustaining innovation in knowledge-intensive sectors (Shane, 2008). These directions promise to build on the current findings, addressing current gaps and fostering a deeper understanding of AI's transformative potential. Future work could also investigate barriers to AI adoption in Indian MSMEs, such as those identified in recent studies on resource constraints and skill gaps (Chatterjee et al., 2024).

## 7. Conclusion

This study examines the structural impact of AI on India's knowledge-intensive startup ecosystem, revealing scaled growth but efficiency challenges through its integration into operational and strategic frameworks (Brynjolfsson & McAfee, 2014). Grounded in GPT theory, it provides practical guidance for entrepreneurs seeking to optimize resources and for policymakers aiming to enhance competitiveness through targeted support (Bresnahan & Trajtenberg, 1995). The findings highlight a shift toward larger structures in AI-era firms, influencing investment strategies and sectoral growth. However, the absence of survival data underscores the need for further research into the long-term sustainability and scalability of these startups as the ecosystem evolves. This investigation lays a foundation for longitudinal studies to assess the enduring effects of AI, offering a roadmap for leveraging technology in emerging markets. As the knowledge-intensive sector continues to grow, these insights shape the future of India's entrepreneurial landscape, contributing to global discourse on innovation and economic development (Schwab, 2017). With India's GenAI startup base growing 3.6x to over 240 in 2024 and funding surpassing $750 million, the trajectory points to sustained AI-driven growth (NASSCOM, 2024).

**Acknowledgements**

**Statements and Declarations**

**Author Contributions:**

Both the authors contributed equally at all stages of the development leading up to the final submission.

**Funding Statement:**

This research did not receive any specific grant from funding agencies in the public, commercial, or not-for-profit sectors.

**Conflict of Interest:** The author declares no conflict of interest related to this research.

**Data Availability:**

The data that support the findings of this study are available from a proprietary startup database, but restrictions apply to the availability of these data, which were used under license for the current study and so are not publicly available. The data are, however, available from the authors upon reasonable request and with the permission of proprietary startup database firm.

**Software & AI Usage Statement:**

The author has made use of JASP for the statistical analyses, ChatGPT 4O and Grammarly to correct the language and the overall writing style of the manuscript. After using these tools, the author reviewed and edited the content as needed and take(s) full responsibility for the content of the published article.